\documentclass[aps,prb,citeautoscript,twocolumn,superscriptaddress,footinbib,10pt]{revtex4-1} 
\usepackage[latin1]{inputenc}
\usepackage{amsmath}
\usepackage{amssymb}
\usepackage{mathrsfs} 
\usepackage{color}
\usepackage{graphicx,color,colortbl}
\usepackage{rotating,array,tabularx,booktabs}
\usepackage{varwidth,xcolor}

\newcommand{\dif}{\mathrm{d}}%

\newcommand{\ii}{\mathrm{i}}%
\newcommand{\rs}{\vec{r}\hskip1pt'}%
\newcommand{\INTR}{\int_{\mathbb{R}^2}\!\!\!\!}%
\newcommand{\INTphi}{\int_0^{2\pi}\!\!\!\!\!\!\!}%

\begin{document}

\title{Liquid crystals of hard rectangles on flat and cylindrical manifolds}
\author{Christoph E. Sitta}
\affiliation{Institut f{\"u}r Theoretische Physik II: Weiche Materie,
Heinrich-Heine-Universit{\"a}t D{\"u}sseldorf, D-40225 D{\"u}sseldorf, Germany}

\author{Frank Smallenburg}
\affiliation{Institut f{\"u}r Theoretische Physik II: Weiche Materie,
Heinrich-Heine-Universit{\"a}t D{\"u}sseldorf, D-40225 D{\"u}sseldorf, Germany}

\author{Raphael Wittkowski}
\affiliation{Institut f\"{u}r Theoretische Physik, Westf\"alische Wilhelms-Universit\"{a}t M\"{u}nster, D-48149 M\"{u}nster, Germany}
\affiliation{Center for Nonlinear Science (CeNoS), Westf\"alische Wilhelms-Universit\"{a}t M\"{u}nster, D-48149 M\"{u}nster, Germany}

\author{Hartmut L\"owen}
\affiliation{Institut f{\"u}r Theoretische Physik II: Weiche Materie,
Heinrich-Heine-Universit{\"a}t D{\"u}sseldorf, D-40225 D{\"u}sseldorf, Germany}

\begin{abstract}
Using the classical density functional theory of freezing and Monte Carlo computer simulations, we explore the liquid-crystalline phase behavior of hard rectangles on flat and cylindrical manifolds. Moreover, we study the effect of a static external field which couples to the rectangles' orientations, aligning them towards a preferred direction. In the flat and field-free case, the bulk phase diagram involves stable isotropic, nematic, tetratic, and smectic phases depending on the aspect ratio and number density of the particles. The external field shifts the transition lines significantly and generates a binematic phase at the expense of the tetratic phase. On a cylindrical manifold, we observe tilted smectic-like order, as obtained by wrapping a smectic layer around a cylinder. We find in general good agreement between our density functional calculations and particle-resolved computer simulations and mention possible setups to verify our predictions in experiments.
\end{abstract}

\maketitle

\section{Introduction}
There are many ways to control structural ordering and topological defects in liquid crystals. One way is to expose them to an external field which aligns the particle orientations and therefore favors the formation of orientationally ordered phases \cite{Hanus1969, WojtowiczS1974, KhokhlovS1982, NicastroK1984, Hornreich1985, LelidisND1993, TangF1993, VargaJS1998, GrafL1999, KimuraKTHIH2005}. Another way is to confine liquid crystals on a curved manifold which enforces the formation of defects due to topological constraints \cite{DzubiellaSL2000, Stark2001, BauschBCDHNNTW2003, FernandezNievesVULMNW2007, ShinBX2008, BatesSZ2010, ArakiBBT2011, LiangSRL2011, LopezLeonKDVFN2011, LiangNZRL2013, MartinezRLVZS2014}. Our fundamental understanding of the structuring of liquid crystals has been strongly aided by simulation studies of hard anisotropic particles, the minimal model required to obtain liquid-crystalline phase behavior \cite{VroegeL1992, BolhuisF1997, vanRoijBMF1995, DonevBST2006, TriplettF2008, GengS2009, WensinkLMHWZKM2013, Dijkstra2014, OettelKDESHG2016}. Moreover, dating back to the seminal work of Onsager \cite{Onsager1949}, these systems have been studied extensively by the density functional theory of freezing (DFT) \cite{PoniewierskiH1988, vanRoijBMF1995, SchlackenMS1998, MartinezRatonVM2005, MartinezRatonVM2006,  HansenGoosM2009, MartinezRatonV2009, HansenGoosM2010, BelliDvR2012, MarechalZL2012, delasHerasMRMV2013, MarechalL2013, WittmannMM2015, OettelKDESHG2016, DiazDeArmasMR2017, MarechalDD2017}. 

Although most studies on liquid crystals of hard particles are performed in three spatial dimensions, two-dimensional systems have been considered extensively as well \cite{SchlackenMS1998, BatesF2000, MartinezRatonVM2005, DonevBST2006, MartinezRatonVM2006, Vink2007, TriplettF2008, MartinezRatonV2009, BelliDvR2012, delasHerasMRMV2013, OettelKDESHG2016}. In two dimensions, the phase behavior is often more subtle: Even in the isotropic limit of hard disks, the crystallization transition is much more complex \cite{KapferK2015, ThorneyworkAAD2017} than in the three-dimensional counterpart being hard-sphere freezing. An additional source of complexity can be formed by the shape of the substrate supporting the particles. For example, two-dimensional sheets of liquid crystals can be constrained on a curved manifold resulting in liquid-crystalline shells \cite{KnorowskiT2012, XingSBYJL2012} with many novel structural ordering phenomena \cite{CoursaultZCBPLBIdMAGGGL2016, ZhangZRdP2016}. These structures can be further tuned by aligning fields to, e.g., control the number of defects in the liquid-crystalline structure \cite{SkavcejZ2008}. One of the simplest substrate shapes is a cylindrical manifold, where one of the principal curvatures vanishes. Interestingly, systematic studies for highly ordered liquid-crystalline phases (like smectics) on cylinders are not available for freely orientable rods or rectangles. Previous work addressed liquid crystals confined between two planar walls (see, e.g., Ref.~\onlinecite{GeigenfeindRSdlH2015}) and the anisotropic dynamics of isotropic disks \cite{MughalW2014, KustersPS2015} or parallel cylinders \cite{MartinezRatonV2013} on a cylindrical manifold.

In this article we combine two aspects of controlling liquid-crystalline ordering, namely aligning external fields and constraints by curved manifolds. We do this for a two-dimensional system of hard rectangles and first study its bulk phase behavior in the flat and field-free case as a function of the particles' aspect ratio and number density. To tackle this problem, we propose a new DFT and perform complementary Monte Carlo (MC) computer simulations, showing stable isotropic \cite{MartinezRatonVM2005, Vink2007, SittaSWL2016}, nematic \cite{MartinezRatonVM2005, MartinezRatonVM2006, NarayanMR2006, Vink2007, delasHerasMRMV2013, DiazDeArmasMR2017}, tetratic \cite{SchlackenMS1998, MartinezRatonVM2005, DonevBST2006, MartinezRatonVM2006, NarayanMR2006, ZhaoHHRC2007, TriplettF2008, GengS2009, MartinezRatonV2009, delasHerasMRMV2013, SanchezAM2016, DiazDeArmasMR2017}, and smectic \cite{MartinezRatonVM2005, NarayanMR2006, BelliDvR2012, delasHerasMRMV2013} phases. Upon applying an aligning external field, the phase transition lines are shifted significantly and a binematic phase occurs at the expense of the tetratic phase. We then consider the same system on a cylindrical manifold, adjusting the external field to favor orientation along the cylindrical perimeter. Interestingly, a new tilted smectic-like order is observed, which emerges from wrapping a smectic layer around a cylinder. This is similar in spirit to helical hard-sphere configurations inside cylinders \cite{MughalCW2011, OguzML2011}. In general, we find good agreement between our DFT calculations and the particle-resolved computer simulations. Our predictions can be verified in real-space experiments on strongly confined colloidal rods \cite{LinCPSWLY2000, GalanisNLH2010, HermesVLVvODvB2011, ZvyagolskayaAB2011, BesselingHKdNDDIvB2015, MuellerdlHRH2015, GuillamatIMS2016, WalshM2016} and granulates \cite{NarayanMR2006, ZhaoHHRC2007, CruzHidalgoZMP2010, HernandezNavarroIMST2012, SanchezAM2016}.

The paper is organized as follows: In Sec.~\ref{chap:methods}, we present our new DFT and describe the MC simulations. The results from the DFT calculations and MC simulations are presented in Sec.~\ref{chap:results}, including the phase diagrams for the flat bulk system of hard rectangles with and without an external field as well as the extension towards a cylindrical manifold. Finally, we conclude in Sec.~\ref{chap:conclusion}.

\section{\label{chap:methods}Methods}
\subsection{Density functional theory}
DFT \cite{Evans1979, Evans1992, Roth2010} provides a versatile framework for determining the equilibrium density profile $\rho_\mathrm{eq}(\vec{r}, \phi)$ of a system of interacting particles. The one-particle density profile $\rho(\vec{r}, \phi)$ represents the probability density for finding a particle with orientation $\phi$ at position $\vec{r}=(x,y)^{\mathrm{T}}$. 
For the rectangular particles considered in this work, $\phi$ denotes the angle measured counterclockwise from the positive $x$ axis to the long axis of a particle. 
The key expression in DFT is the grand-canonical free-energy functional
\begin{equation}
\Omega[\rho(\vec{r}, \phi)] = \mathcal{F}[\rho(\vec{r}, \phi)] 
+ \INTR \dif^2 r \INTphi \dif \phi \, \rho(\vec{r}, \phi) (V_{\mathrm{ext}}(\vec{r},\phi) - \mu)\,,
\label{eq:GP_functional}%
\end{equation}
which is minimized by $\rho_\mathrm{eq}(\vec{r}, \phi)$ and whose value $\Omega[\rho_\mathrm{eq}(\vec{r}, \phi)]$ matches the equilibrium value of the grand potential of the system. In Eq.\ \eqref{eq:GP_functional}, $\mathcal{F}[\rho(\vec{r}, \phi)]$ is the Helmholtz free-energy functional, $V_\mathrm{ext}(\vec{r},\phi)$ is the external potential acting on the particles, and $\mu$ is the chemical potential. Unfortunately, $\mathcal{F}[\rho(\vec{r}, \phi)]$ is rarely known except for a few special cases (e.g., for monodisperse hard particles in one spatial dimension \cite{Percus1976}) and therefore usually needs to be approximated. It is commonly split as 
$\mathcal{F} = \mathcal{F}_{\mathrm{id}} + \mathcal{F}_{\mathrm{exc}}$
into a sum of the analytically known ideal-gas contribution $\mathcal{F}_{\mathrm{id}}$ and an unknown excess term $\mathcal{F}_{\mathrm{exc}}$. The ideal-gas term is given by
\begin{equation}
\beta \mathcal{F}_{\mathrm{id}} = \INTR \dif^2 r \INTphi \dif\phi 
\,\rho(\vec{r}, \phi) \big(\ln(2 \pi \Lambda^2 \rho(\vec{r}, \phi)) - 1 \big)
\end{equation}
with the inverse thermal energy $\beta$ (also called thermodynamic beta) and the thermal de Broglie wavelength $\Lambda$.

For the excess term we propose a phenomenological approximation   
\begin{equation}
\beta \mathcal{F}_{\mathrm{exc}} = \INTR \dif^2 r \,  \Phi(\vec{r})
\end{equation}
with the rescaled excess free-energy density $\Phi(\vec{r})$, which features all phases that were observed by simulations \cite{DonevBST2006, MartinezRatonVM2006, TriplettF2008, GengS2009} and experiments \cite{NarayanMR2006, ZhaoHHRC2007, SanchezAM2016} for a fluid of hard rectangles, i.e., an isotropic, a nematic, a tetratic, and a smectic phase. An illustration of the different phases is given by Fig.\ \ref{fig:1} further below.
We write $\Phi(\vec{r})$ as a sum of four terms,
\begin{equation}
\label{eq:energy_density}
\Phi(\vec{r}) = \Phi_\mathrm{Ons}(\vec{r}) + \Phi_\mathrm{FMT}(\vec{r}) + \Phi_\mathrm{T}(\vec{r}) + \Phi_\mathrm{P}(\vec{r})\,,
\end{equation}
which are explained and justified in detail in the following.

To equip our functional with an appropriate low-density limit, we start by using the Onsager approximation \cite{Onsager1949} 
\begin{equation}
\Phi_\mathrm{Ons}(\vec{r}) = - \frac{1}{2} \INTphi \dif\phi \,\rho(\vec{r}, \phi)\,\INTR \dif^2 r' \INTphi \dif \phi' \rho(\rs, \phi') f(\vec{r} - \rs, \phi, \phi') 
\label{eq:ONS}
\end{equation}
with the Mayer function
\begin{equation}
f(\vec{r} - \rs,\phi,\phi') =
\begin{cases} %
-1 \;, &  \text{if particles with coordinates }\\
      & (\vec{r}, \phi) \text{ and } (\rs,\phi') \text{ overlap,}   \\
0 \;, &  \text{otherwise,} 
\end{cases}%
\label{eq:MayerFunction}%
\end{equation}
which should perform well in the low-density limit of long rods. As this term scales with second order in density, our functional $\mathcal{F}_{\mathrm{exc}}$ will inherit this low-density behavior as long as all other terms on the right-hand side of Eq.\ \eqref{eq:energy_density} scale with third or higher order in density. The term $\Phi_{\mathrm{Ons}}(\vec{r})$ already yields an isotropic and a nematic phase but does not feature spatial density modulations, which are necessary, e.g., for a smectic phase.

Thus, another term is needed to allow a smectic phase at high densities. As the particles should be well aligned at high densities, we adopt the excess free-energy functional for parallel hard rectangles, which was derived by Cuesta and {Mart{\'{\i}}nez-Rat{\'o}n} by a dimensional crossover \cite{CuestaMR1997b, CuestaMR1997}, and therefore introduce the term 
\begin{equation}
\Phi_\mathrm{FMT}(\vec{r}) = n_2(\vec{r}) \Big( -n_0(\vec{r}) \ln\!\big(1-n_2(\vec{r})\big) + \frac{n_{1x}(\vec{r}) \, n_{1y}(\vec{r})}{1-n_2(\vec{r})} \Big)
\label{eq:FMT}%
\end{equation}
in our rescaled free-energy density. Here, the $n_i(\vec{r}) $ with $i\in\{0,1x,1y,2\}$ are weighted densities as typical for fundamental measure theory \cite{Rosenfeld1989, Roth2010}. In the case of freely orientable rectangles, they are defined as in Ref.\ \onlinecite{MartinezRatonVM2005} by the angle-integrated cross-correlations
\begin{equation}
\begin{split}%
n_i(\vec{r}) &= \INTphi \dif \phi \, [\rho \star \omega^{(i)}](\vec{r}, \phi) \\
&= \INTphi \dif \phi \INTR \dif^2 r' \, \rho(\rs, \phi) \omega^{(i)}(\rs - \vec{r}, \phi)
\end{split}%
\end{equation}
of the one-particle density $\rho(\vec{r}, \phi)$ with the geometric weight functions
{\allowdisplaybreaks\begin{align}%
\omega^{(0)}(\vec{r}, \phi) &= \frac{1}{4} \delta\Big(\frac{D}{2} 
- |x_{\phi}|\Big) \delta\Big(\frac{L}{2} - |y_{\phi}|\Big)\,,\\
\omega^{(1x)}(\vec{r}, \phi) &= \frac{1}{2} \delta\Big(\frac{D}{2} - |x_{\phi}|\Big) \Theta\Big(\frac{L}{2} - |y_{\phi}|\Big)\,,\\
\omega^{(1y)}(\vec{r}, \phi) &= \frac{1}{2} \Theta\Big(\frac{D}{2} - |x_{\phi}|\Big) \delta\Big(\frac{L}{2} - |y_{\phi}|\Big)\,,\\
\omega^{(2)}(\vec{r}, \phi) &= \Theta\Big(\frac{D}{2} - |x_{\phi}|\Big) \Theta\Big(\frac{L}{2} - |y_{\phi}|\Big)\,.
\end{align}}%
Here, $\delta(x)$ is the Dirac delta function and $\Theta(x)$ is the Heaviside function; $D$ and $L$ with $D<L$ denote the width and length of the rectangular particles, respectively, and $x_{\phi}$ and $y_{\phi}$ are defined as $x_{\phi}=x\cos(\phi)-y\sin(\phi)$ and $y_{\phi}=x\sin(\phi)+y\cos(\phi)$.
Except for the prefactor $n_2(\vec{r})$, Eq.\ \eqref{eq:FMT} is identical to the corresponding expression from Cuesta and {Mart{\'{\i}}nez-Rat{\'o}n} \cite{CuestaMR1997b, CuestaMR1997}. As their original expression scales with $\mathcal{O}(\rho^2)$, we need the dimensionless prefactor $n_2(\vec{r})$, which scales with order $\mathcal{O}(\rho)$, in Eq.\ \eqref{eq:FMT} to maintain the low-density behavior from the Onsager term in Eq.\ \eqref{eq:energy_density}.

Since the two previous terms do not feature a stable tetratic phase yet, we introduce an empirical term $\Phi_{\mathrm{T}}(\vec{r})$, which suppresses nematic order and favors tetratic order. 
We make use of the squared Fourier coefficients $|A_2(\vec{r})|^2$ and $|A_4(\vec{r})|^2$, which are defined as
\begin{equation}
|A_n(\vec{r})|^2 =\frac{1}{\bar{\rho}(\vec{r})^2} \bigg( \INTphi \dif\phi \, \rho(\vec{r},\phi) \, e^{-\ii n \phi}\bigg) \bigg( \INTphi \dif\phi \,  \rho(\vec{r},\phi) \, e^{\ii n \phi}\bigg)
\end{equation}
with the angle-integrated center-of-mass density (i.e., concentration field) $\bar{\rho}(\vec{r}) =  \int_{0}^{2\pi} \!\dif\phi\, \rho(\vec{r},\phi)$. 
On the one hand, $|A_2(\vec{r})|^2 \to 1$ and  $|A_4(\vec{r})|^2 \to 1$ holds for twofold nematic order in a perfect nematic phase. On the other hand, $|A_2(\vec{r})|^2 \to 0$ and $|A_4(\vec{r})|^2 \to 1$ holds for fourfold tetratic order in a perfect tetratic phase. 
Thus a term as simple as $\Phi_{\mathrm{T}} \propto a \, |A_2|^2 - |A_4|^2$ will favor tetratic order while suppressing nematic order if $a>0$, whereas $a<0$ will favor nematic order. Since tetratic order is present only in the tetratic phase, while the nematic and smectic phases include nematic order, we require the switching coefficient $a = a(L/D, n_{2})$ to dependent on both the aspect ratio $L/D$ and the weighted density $n_{2}(\vec{r})$ in order to adjust the phases at the correct positions in the phase diagram. Choosing $a = \chi_{\mathrm{T}}(L/D) - n_2(\vec{r})$ allows $a$ to switch between negative values for local area fractions $n_2(\vec{r})$ above an aspect-ratio-dependent threshold area fraction $\chi_{\mathrm{T}}(L/D)> 0$ and positive values below this threshold. For the area fraction $\chi_{\mathrm{T}}(L/D)$ describing the transition between nematic and tetratic states, we shift a $\tanh$ to positive values by writing $\chi_{\mathrm{T}}(L/D) = 0.5 c_1 \big( 1 + \tanh(c_2(c_3 - L/D)) \big)$ and use fitted values for its amplitude ($c_1 = 0.85$), steepness ($c_2 = 2/3)$, and turning point ($c_3 = 9$) to match our MC simulation data. 
The full tetratic term then reads
\begin{equation}
\begin{split}
\Phi_{\mathrm{T}} = \frac{5}{2} \, \bar{\rho}(\vec{r}) \, n_2(\vec{r})^2 \, \Big( 1 &+ \big(\chi_{\mathrm{T}}(L/D) - n_2(\vec{r})\big) |A_2|^2 \\
&- \chi_{\mathrm{T}}(L/D) |A_4|^2 \Big)
\label{eq:tetratic}
\end{split}
\end{equation}
with the threshold area fraction
\begin{equation}
\chi_{\mathrm{T}}(L/D) = 0.425 \, \Big( 1 + \tanh\!\big(\tfrac{2}{3} \, (9 - L/D)\big) \Big) \,.
\end{equation}
In Eq.\ \eqref{eq:tetratic}, the summand $1$ in the outer parentheses and the positive prefactor $\chi_{\mathrm{T}}(L/D)$ in front of $|A_4|^2$ are included to improve numerical stability. The prefactor $(5/2) \bar{\rho}(\vec{r}) \, n_2(\vec{r})^2$ with the angle-integrated local density $\bar{\rho}(\vec{r})$ ensures both that the dimensions of $\Phi_{\mathrm{T}}$ are correct and that $\Phi_{\mathrm{T}}$ scales with $\mathcal{O}(\rho^3)$ so that the low-density behavior of the Onsager term is maintained in Eq.\ \eqref{eq:energy_density}. In this prefactor, the proportionality constant $5/2$ is chosen to match the MC simulation data and constitutes the fourth fitted parameter in this model.

Finally, we add the penalty term
\begin{equation}
\Phi_\mathrm{P}(\vec{r}) = -\frac{n_2(\vec{r})^2}{D^2}  \ln\!\big(1-n_2(\vec{r})\big)\,,
\label{eq:penalty}
\end{equation}
which diverges for local area fractions $n_2(\vec{r}) \to 1$ and avoids an unphysical overlap of the hard particles. Again, a prefactor $(n_2(\vec{r})/D)^2$ is chosen for reasons of dimensionality and to maintain the low-density limit described by the Onsager term in Eq.\ \eqref{eq:energy_density} through $\Phi_{\mathrm{P}}(\vec{r})$ scaling with $\mathcal{O}(\rho^3)$.

After inserting Eqs.\ \eqref{eq:ONS}, \eqref{eq:FMT}, \eqref{eq:tetratic}, and \eqref{eq:penalty} into the rescaled excess free-energy density \eqref{eq:energy_density} and choosing an expression for the external potential $V_{\mathrm{ext}}(\vec{r},\phi)$, the equilibrium density $\rho_\mathrm{eq}(\vec{r}, \phi)$ can be obtained by a free minimization of the functional \eqref{eq:GP_functional} with respect to $\rho(\vec{r}, \phi)$.
When an aligning external field is taken into account, we choose the external potential as
\begin{equation}
V_{\mathrm{ext}}(\vec{r},\phi) = V_{0} \sin(\phi)^{2} 
\label{eq:ext_potential}%
\end{equation}
with the amplitude $V_0$. Otherwise, $V_{\mathrm{ext}}(\vec{r},\phi)$ is set to zero. Note that these expressions for $V_{\mathrm{ext}}$ maintain the $\phi\to\phi +\pi$ invariance of the system. 

As in Refs.\ \onlinecite{SittaSWL2016,WittmannSSL2017}, the minimization of the functional \eqref{eq:GP_functional} is performed numerically in real space by using a Picard iteration scheme \cite{Roth2010}
\begin{equation}
\begin{split}%
\rho^{(i+1)}(\vec{r}, \phi) &= (1-{\alpha}) \rho^{(i)}(\vec{r}, \phi) \\
&\;\, + \frac{\alpha}{2\pi\Lambda^2}\exp{\!\bigg(\beta \Big(\mu^{(i)} - V_{\mathrm{ext}}(\vec{r},\phi) - \frac{\delta\mathcal{F}_{\mathrm{exc}}}{\delta\rho(\vec{r},\phi)} \Big)\! \bigg)}
\end{split}\raisetag{9ex}%
\end{equation}
with the mixing parameter ${\alpha} \le 0.01$, $\Lambda$ set to $D/\sqrt{2\pi}$, and the functional derivative $\delta\mathcal{F}_{\mathrm{exc}}/\delta\rho(\vec{r},\phi)$. To maintain a constant area fraction, the chemical potential $\mu^{(i)}$ is recalculated in every iteration step $i$. It converges to a finite value during the iteration. As in previous works \cite{OettelDBNS2012,HaertelOREHL2012, SittaSWL2016, WittmannSSL2017}, we combine this iteration with a direct inversion in the iterative subspace \cite{Ng1974,Pulay1980,Pulay1982,KovalenkoTNH1999} to improve the convergence.
The orientations of the rectangles are discretized in equidistant steps of $\Delta \phi \le \pi/24$ and a spatial Cartesian grid with step sizes $\Delta x = \Delta y \approx 0.03 D$ is used.
For the simulation box, a rectangular domain with a size much larger than that of a particle and with periodic boundary conditions is used. When considering a flat system, we minimize the grand-canonical free energy per area also with respect to the width and length of the simulation box. 
In the case of particles on a cylindrical manifold, the width of the box is kept constant and equal to the circumference of the cylinder, while its length is varied. 
In both cases, the width and length of the box shall correspond to the $x$ and $y$ directions of our Cartesian coordinate system, respectively.

\subsection{\label{sec:MC}Monte Carlo simulations}
In order to estimate the bulk phase behavior of hard rectangles in the regime of interest, we make use of MC simulations. In particular, we simulate perfectly hard rectangular particles in rectangular boxes with periodic boundary conditions, at constant number of particles $N$, pressure $P$, and temperature $T\propto 1/\beta$. Overlaps between rectangles are detected using the separating axis theorem (see, e.g., Ref.\ \onlinecite{GottschalkLM1996}). Simulations consist of single-particle translations and rotations, as well as cluster movements that collectively rotate all particles whose centers lie in a small circular region around the center of a random particle by 90 degrees. We estimate the isotropic-to-nematic and isotropic-to-tetratic transitions by measuring the average nematic and tetratic order parameters in the system, which are defined as
\begin{equation}
S_k = \left | \frac{1}{N} \sum_{j=1}^N \exp(\ii k \phi_j) \right|^{2} 
\end{equation}
with $k=2$ for the nematic and $k=4$ for the tetratic order parameter. Here, $\phi_j$ is the angle measured from the $x$ axis of the system to the long axis of the $j$th particle. These order parameters are zero for an isotropic system. In a perfectly nematic phase, where all particles are aligned along one axis, $S_2 = 1$ and $S_4 = 1$, while in a perfectly tetratic phase $S_2 = 0$ and $S_4 = 1$. Since the boundary between the smectic phase and the lower-density phases is a first-order phase transition, hysteresis makes exact identification of this transition difficult. To estimate the melting line for the smectic phase, we start simulations in the smectic phase and determine at which density the layering disappears by visual inspection. Note that while we refer to this phase as smectic in this work, we did not closely examine the decay of translational ordering in the system and hence do not resolve any distinction between a crystalline and a smectic phase, which would both show similar layering.

We follow the same approach for determining the phase diagram for rectangles in an aligning field, where we apply the external potential \eqref{eq:ext_potential}. 
To explore self-assembly on a cylindrical surface, we fix the width of the periodic simulation box to the desired circumference $C$ of the cylinder and allow volume fluctuations only along the perpendicular direction. The length of the simulation box is always much larger than $C$. To compare more directly to the DFT results with constant area fraction, we first compress the system from a low-density isotropic fluid to the desired density by slowly ramping up the pressure, and then fix the volume once the desired volume is reached. In this effectively one-dimensional system, there are no true phase transitions. As a result, the system typically fluctuates between qualitatively different structures with respect to time or with respect to the position in the simulation box. Hence, we usually find a variety of likely states for a given combination of the cylinder circumference in units of particle length $C/L$, the aspect ratio $L/D$ of the particles, their total area fraction $\eta$, and the amplitude $V_0$ of the external potential. To address this ambiguity, we perform multiple independent simulation runs at each state point and collect data on the observed structures by visual inspection.
The simulations involve $N=1000$-$4000$ particles for the flat space and $N=120$ particles on a cylindrical surface.

\section{\label{chap:results}Results}
In this section, we explore the self-assembly of hard rectangles. We first test the developed functional on flat systems without an aligning external field and then apply it to both flat systems with an external field  and to systems on a cylindrical manifold.

\subsection{\label{chap:bulk}Phase behavior on a plane without an external field}
\begin{figure*}[htb]
\centering
\includegraphics[width=\linewidth]{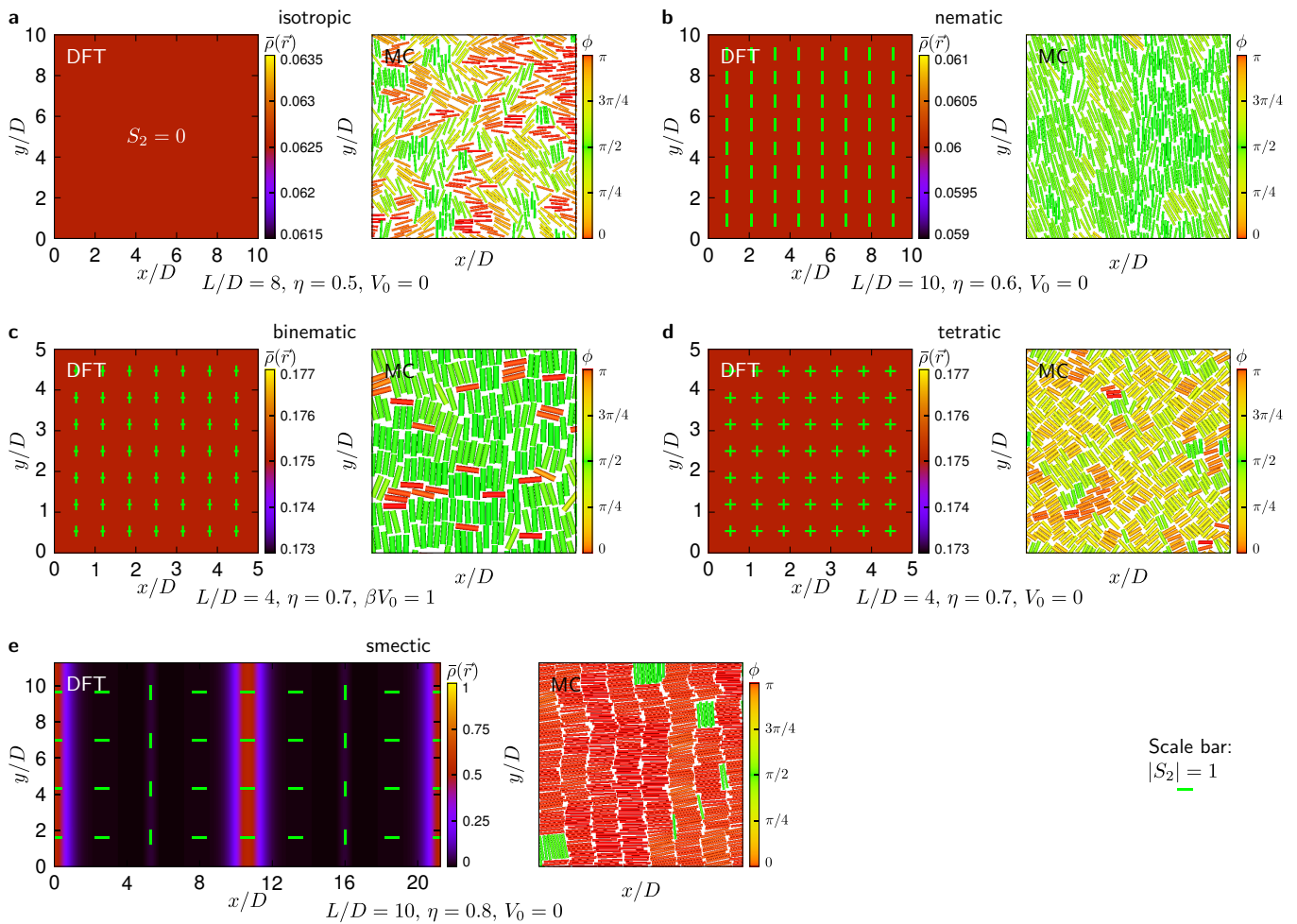}
\caption{\label{fig:1}Typical equilibrium density profiles $\bar{\rho}(\vec{r})$ and orientation fields indicating the local particle alignment (green dashes) obtained from DFT calculations as well as snapshots of MC simulations are shown for all observed types of structures of hard rectangular particles in two spatial dimensions. When $V_0 \neq 0$, an external field that aligns the particles in parallel to the $x$ axis is present. Perfect isotropic and tetratic phases were only found without such an external field, while a binematic phase was only found in the presence of an external field.}%
\end{figure*}

For flat systems of hard rectangles in the absence of any aligning fields, we find four distinct phases in the parameter range considered here: an isotropic phase, a nematic phase, a tetratic phase, and a smectic phase. Typical equilibrated systems of rectangular particles obtained from DFT calculations and MC simulations are shown in Fig.\ \ref{fig:1} for all observed phases. 

In the \textit{isotropic} phase (see Fig.\ \ref{fig:1}a), the particles are disordered with respect to both position and orientation. This phase is observed at low densities for all aspect ratios. Also the \textit{nematic} and \textit{tetratic} phases are disordered in space, but they show an anisotropic distribution of the orientation. We find a \textit{nematic} phase (see Fig.\ \ref{fig:1}b) at intermediate area fractions for large aspect ratios, which are $L/D \gtrsim 7$ in DFT calculations and $L/D \gtrsim 9$ in MC simulations. Although spatially disordered (i.e., spatial correlations are absent in DFT results and decay exponentially in MC simulations), the orientational distribution shows a twofold symmetry, indicating that most particles are aligned parallel to a certain axis. 
On the other hand, the \textit{tetratic} phase (see Fig.\ \ref{fig:1}d) shows a fourfold symmetry in the orientational distribution, which indicates alignment along two perpendicular axes. This phase is found at intermediate area fractions for small aspect ratios. 
For high area fractions, we observe a transition to a spatially ordered \textit{smectic} phase (see Fig.\ \ref{fig:1}e), where aligned particles form layers, with their orientations perpendicular to the layers. This is also known as a smectic A phase. In this phase, we also find particles that are located between and oriented parallel to the layers, which is in agreement with observations in three spatial dimensions \cite{vanRoijBMF1995}.

\begin{figure}[htb]
\centering
\includegraphics[width=\linewidth]{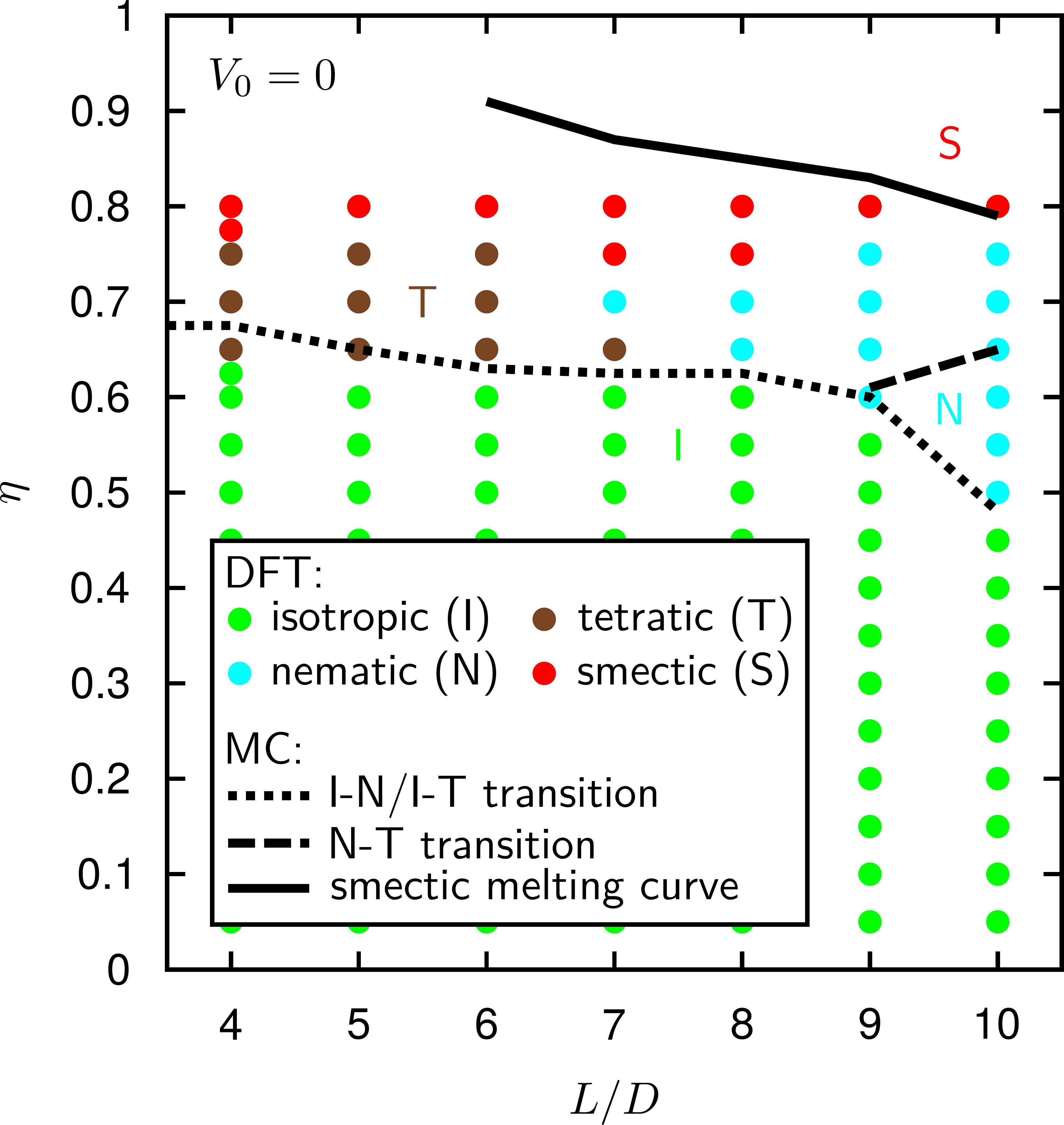}%
\caption{\label{fig:2}Bulk phase diagram of freely orientable hard rectangular particles on a plane without an external field obtained from both DFT calculations and MC simulations. The aspect ratio $L/D$ and area fraction $\eta$ of the particles are varied.}%
\end{figure}

Figure \ref{fig:2} shows the DFT and MC results for the phase diagram of rectangular particles on a plane. 
Both approaches lead to qualitatively similar phase diagrams that include the same phases. Note that, in principle, we expect crystalline phases at very high area fractions for all aspect ratios, including a solid where the particles are tetratically ordered \cite{DonevBST2006} and a periodic crystal of aligned particles on a rectangular lattice. However, it is difficult to distinguish these phases from the tetratic fluid and smectic phases, respectively. Hence, we do not attempt to distinguish between the tetratic solid and fluid or between the orientationally ordered crystal and smectic phase in this work. Instead, we refer to them simply as tetratic and smectic phases, respectively.

\subsection{\label{chap:pot}Phase behavior on a plane with an external field}
\begin{figure}[htb]
\centering
\includegraphics[width=\linewidth]{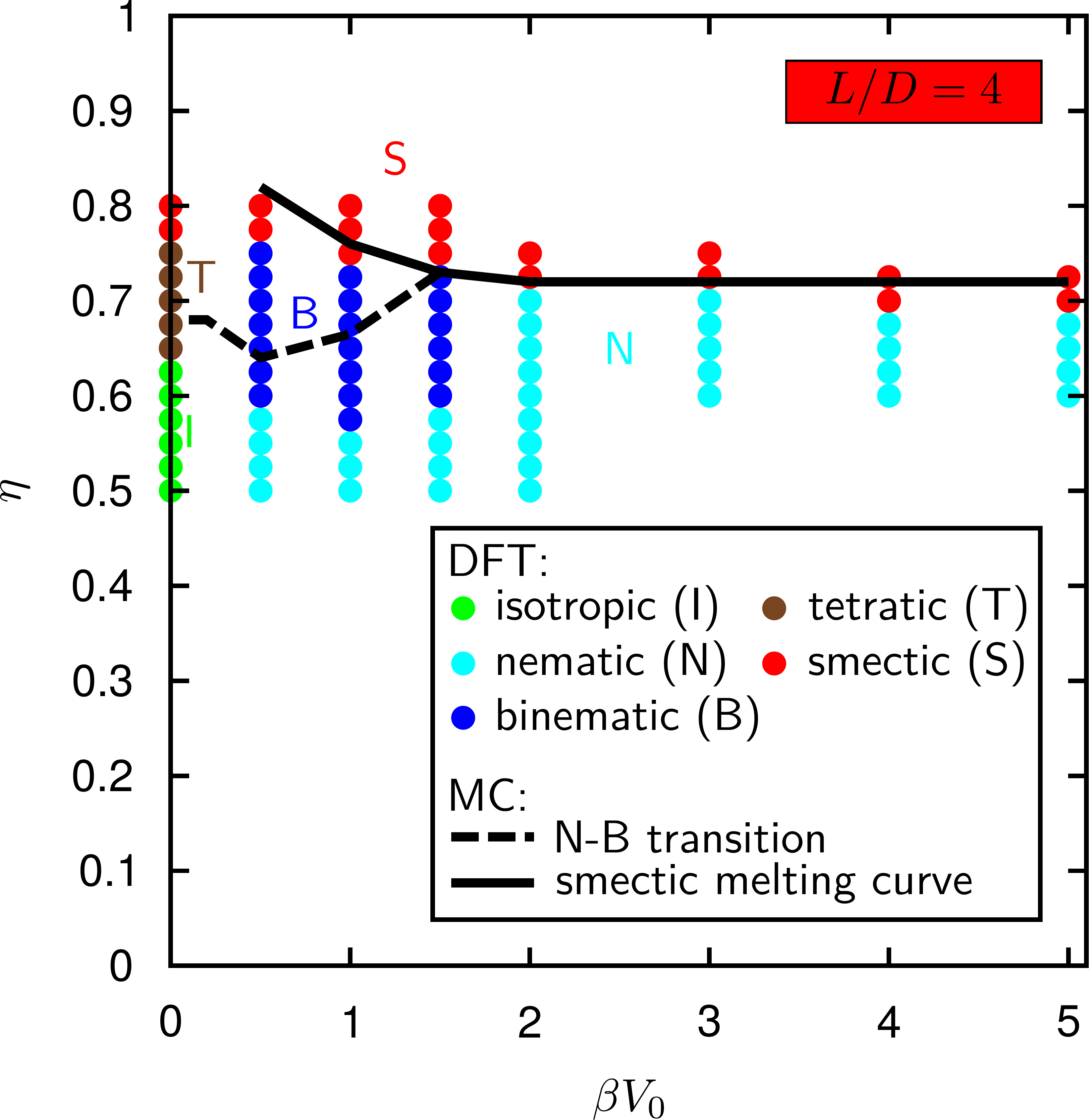}%
\caption{\label{fig:3}Phase diagram of freely orientable hard rectangular particles on a plane in the presence of an aligning external field for both DFT calculations and MC simulations. The aspect ratio of the particles is now $L/D=4$, whereas the amplitude $V_0$ of the external potential and the particles' area fraction $\eta$ are varied. Perfect isotropic or tetratic phases are found only for $V_0=0$.}%
\end{figure}

We now extend our approach to systems of hard rectangles on a plane that are exposed to an aligning external potential (see Eq.\ \eqref{eq:ext_potential}), which acts purely on the orientation of each particle.
To investigate the effect of the potential's amplitude $V_0$ on the phase diagram, we now keep the aspect ratio of the particles fixed at $L/D = 4$ and show the phase diagram for varying potential amplitudes $V_0$ and area fractions $\eta$ in Fig.\ \ref{fig:3}. 
A striking effect of the aligning field is the complete absence of the isotropic and tetratic phases for potential amplitudes $V_0 > 0$. As the field favors alignment of the particles along the $x$ axis, it makes a purely isotropic phase impossible. Likewise, the tetratic phase with four equally pronounced peaks in the orientational distribution is no longer possible, as the probability of alignment along the $x$ axis will always be larger than the probability of alignment along the $y$ axis. When considering, for example, a system with $L/D=4$ and $\eta = 0.7$, the tetratic phase is stable without an external field, but turns smoothly into a nematic phase (with preferred orientation along the $x$ axis) when increasing the potential amplitude $V_0$. During this transition, the peaks in the orientational distribution that correspond to the $y$ direction gradually decrease. As long as the height of the former tetratic peaks is still at least 10\% of the height of the main peaks corresponding to the $x$ direction, we call this intermediate phase \textit{binematic} (see Fig.\ \ref{fig:1}c).

In the phase diagram (see Fig.\ \ref{fig:3}), the results of the DFT calculations and MC simulations show the same trends. When increasing the external field from $V_{0}=0$ to $\beta V_0\le 1$, the binematic phase becomes stable at lower densities for low aspect ratios. Here, the field helps the particles to align parallel or antiparallel to the $x$ axis, but it is too weak to cause full alignment, hence favoring a binematic phase. As the external field is further increased, it eventually causes (nearly) full alignment of the particles and a purely nematic phase becomes stable.
Similarly, increasing the field strength enhances the stability of the smectic phase, where the particles are aligned along one axis. At large field strengths with $\beta V_0\ge 4$ in the case of DFT calculations and $\beta V_0\ge 2$ for MC simulations, this effect saturates and no further stabilization of the smectic phase is observed. In this parameter region, the particles in both the nematic and smectic phases are essentially fully aligned, and hence further increasing the field strength has no effect on the relative stability of the phases. It is important to note that while the parameters of our density functional were chosen in Eq.\ \eqref{eq:tetratic} to improve the agreement with the MC simulation data for systems without an external field, we made no adaptations to the functional for the case with an aligning field. Therefore, it is remarkable that the phase behavior predicted by our DFT calculations and MC simulations still shows good agreement when an aligning external field is present.

\subsection{\label{chap:cyl}Phase behavior on a cylinder with an external field}
We now turn our attention from rectangles on the plane to rectangles on the lateral surface of an infinitely long cylinder, whose axis is parallel to the $y$ axis. To investigate the effect of the periodic confinement on a cylinder, we vary the radius of the cylinder such that its circumference $C$ ranges between $9D = 2.25L$ and $21D = 5.25 L$ for a fixed aspect ratio $L/D=4$ of the rectangles. The lower limit is sufficiently large to avoid cases where two particles could interact with each other on both sides of the cylinder. 
In order to prevent the system from simply forming nematic and smectic phases with the preferred particle orientation parallel to the axis of the cylinder, which would result in a phase diagram quite similar to that for the flat case, we include an external field to align the particles along the $x$ direction. This promises interesting results, because it favors the formation of smectic phases where the particles are aligned along the (short) circumference of the cylinder. Such an alignment results in a competition between the favored layer spacing of the smectic phase and the fixed circumference $C$, leading to more complex self-assembled structures that attempt to satisfy both constraints. 

\begin{figure}[tb]
\centering
\includegraphics[width=\linewidth]{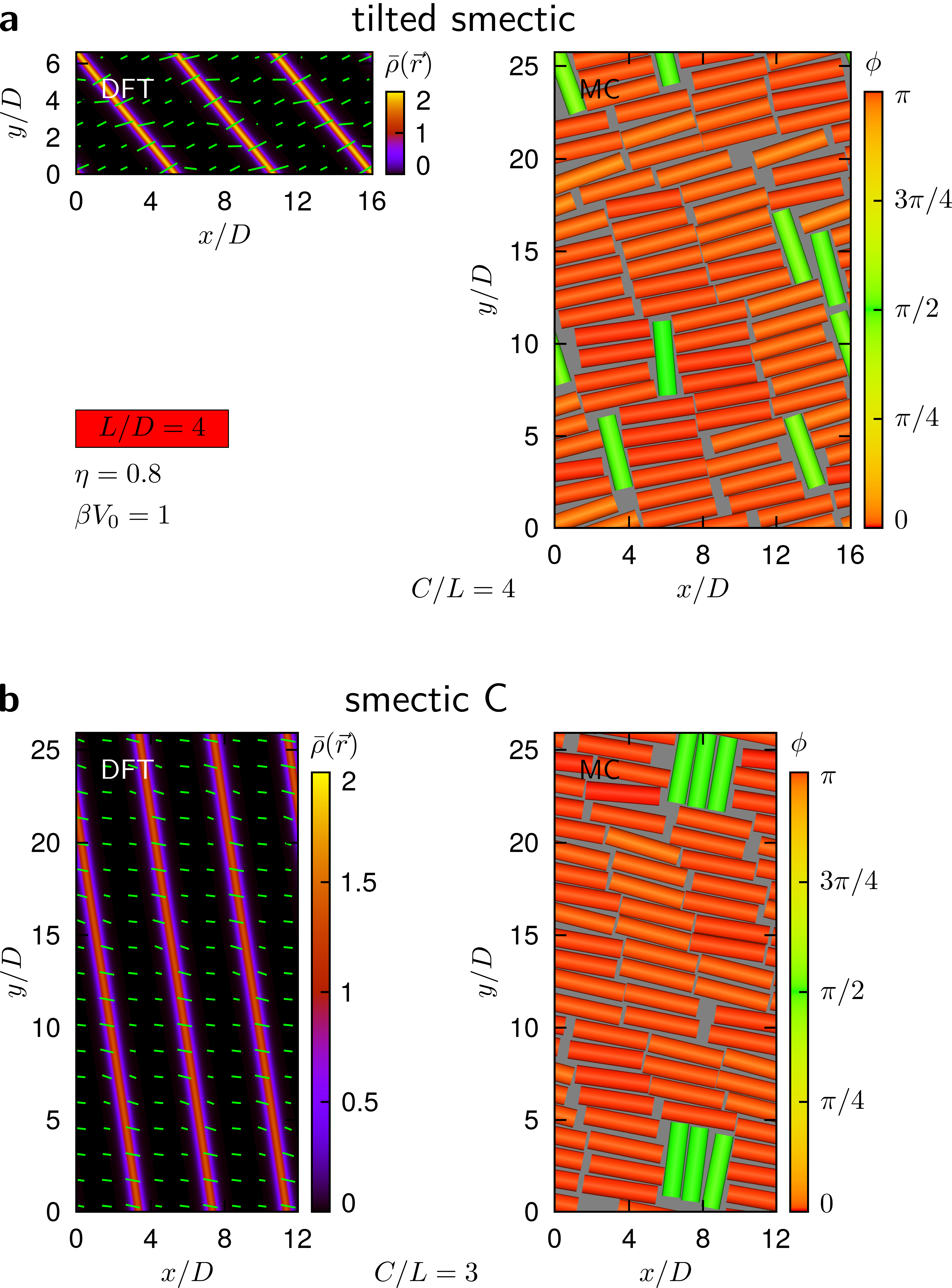}%
\caption{\label{fig:4}As in Fig.\ \ref{fig:1}, but now for a (a) tilted smectic phase and a (b) smectic C phase, which are observed only on a cylinder in the presence of an external field that favors particle alignment along the $x$ direction. The plots show the unrolled cylindrical surface (see Fig.\ \ref{fig:5}b for the snapshots of MC simulations on a cylinder). In both cases, the aspect ratio of the particles is $L/D=4$, their area fraction is $\eta=0.8$, and the amplitude of the external potential is $V_{0}=1/\beta$.}%
\end{figure}

As on a plane, we observe a binematic phase without spatial order and a smectic phase where the particles are aligned according to the external field. In the latter phase, the layers are parallel to the axis of the cylinder. In addition to these phases, we observe two new phases that occur only on a cylinder: firstly, a \textit{tilted smectic} phase with layers along any other direction than the cylinder axis and particle orientations still orthogonal to the layers (see Fig.\ \ref{fig:4}a) and, secondly, a \textit{smectic C} phase, where the particles are no longer oriented perpendicular to the layers (see Fig.\ \ref{fig:4}b). 
We observe these two phases in both DFT calculations and MC simulations. In the MC simulations, two further phases are found: a columnar phase and a tilted columnar phase with particle layers parallel to the particle orientation.

\begin{figure}[htb]
\centering
\includegraphics[width=\linewidth]{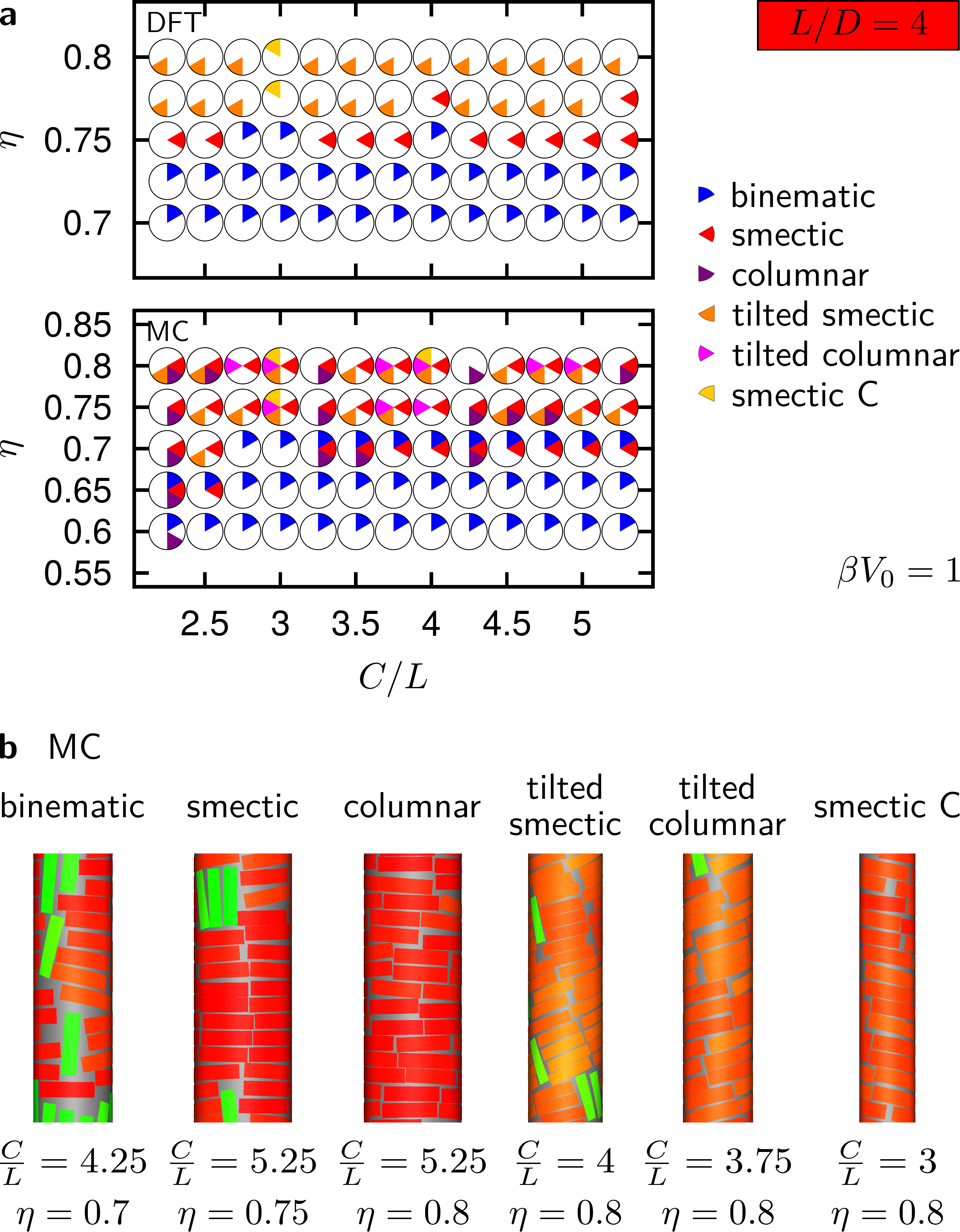}%
\caption{\label{fig:5}(a) Phase diagrams obtained from DFT calculations and MC simulations as well as (b) MC snapshots illustrating the individual phases for hard rectangular particles of aspect ratio $L/D=4$ on a cylinder, where an external field with $\beta V_{0}=1$ is present. We focus here on circumferences $C$ and area fractions $\eta$ close to the transition between spatially ordered and disordered phases.}%
\end{figure}

Figure \ref{fig:5} displays the phase behavior close to the transition between the binematic and the smectic phase obtained from DFT calculations and MC simulations for different cylinder radii.
As one might expect, the smectic phase is most stable for circumferences $C$ just above an integer number of particle lengths $L$ and least stable for circumferences equal to or just below an integer number of particle lengths. These circumferences correspond to cases where an integer number of smectic layers fits, or does not fit, onto the cylinder in the direction preferred by the field, respectively. Although observed at $\eta = 0.75$ on a plane, no inhomogeneous density profiles with smectic layers are found at circumferences $2.75L$, $3L$, and $4L$, when using the DFT. In the MC simulations, we observe multiple competing states for most choices of the circumference and area fraction (see Sec.\ \ref{sec:MC}). Figure \ref{fig:5} provides an overview of structures visually identified from five independent simulation runs at each state point. Note that, since in the finite systems considered here there are no true phase boundaries between different states, classification of different phases is partially subjective.  

\begin{figure}[htb]
\centering
\includegraphics[width=\linewidth]{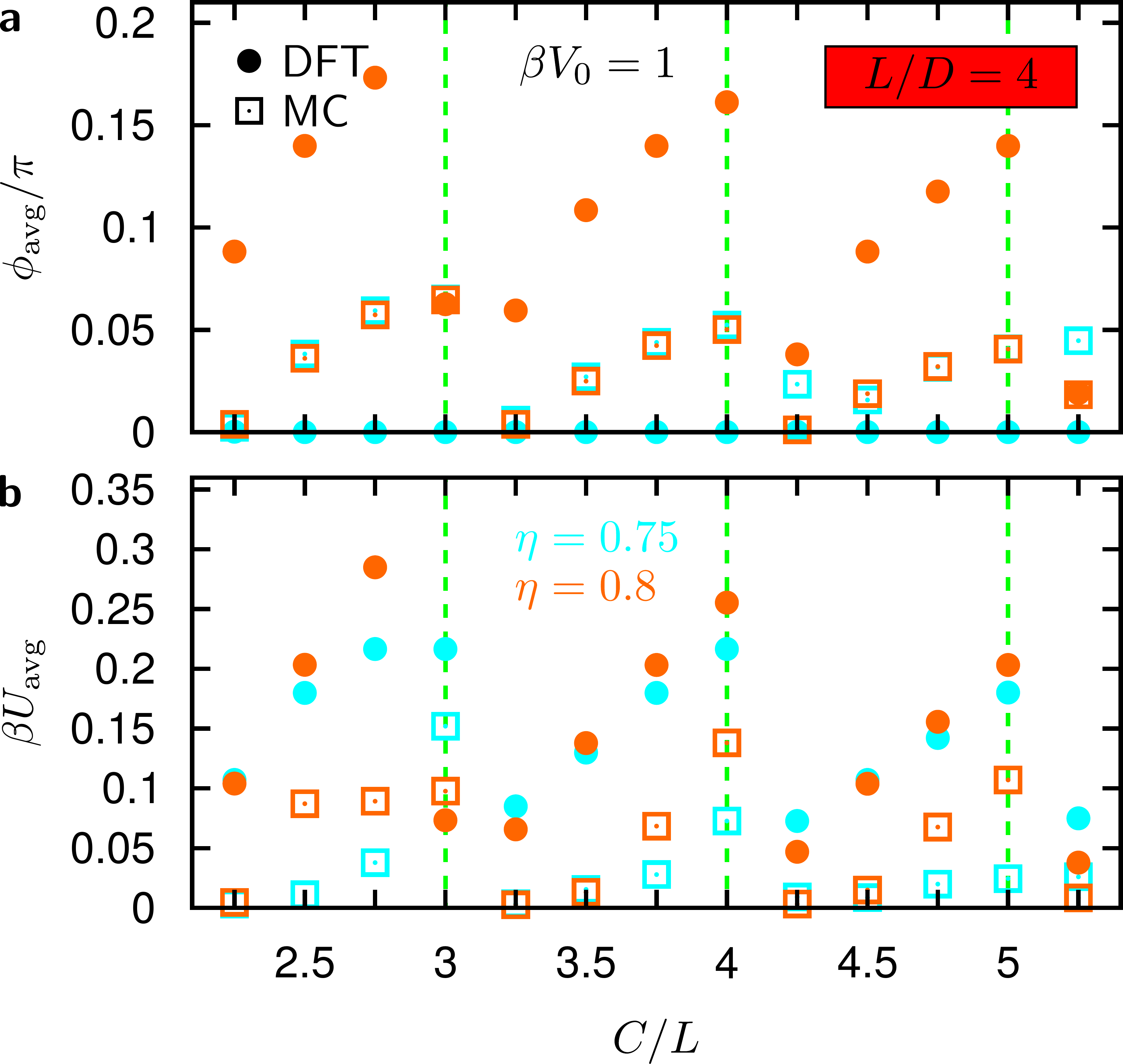}%
\caption{\label{fig:6}(a) The average orientation angle $\phi_{\mathrm{avg}}\in [0,\pi)$ due to the particle alignment in the external field and (b) the average potential energy per particle $U_{\mathrm{avg}}$ in a system of hard rectangular particles with aspect ratio $L/D=4$ on a cylinder are shown as a function of the cylinder circumference $C$ for different area fractions $\eta$ and the potential amplitude $V_{0}=1/\beta$. Circumferences that are exactly an integer multiple of the particle length $L$ are indicated by dashed lines.}%
\end{figure}

Interestingly, for increasing area fractions we observe an increasing tilt of the smectic layers away from the cylinder axis. This is understandable, as at lower area fractions the system can more easily distort or incorporate defects that allow for a better total alignment of the system.
The observed increasing tilt of the particles is further characterized by Fig.\ \ref{fig:6}. There, we show the average particle orientation $\phi_{\mathrm{avg}}\in [0,\pi)$ (see Fig.\ \ref{fig:6}a) relative to the $x$ axis, i.e., to the direction along the circumference, for the area fractions $\eta=0.75$ and $0.8$ and for both the DFT calculations and MC simulations. 
As extracting the average particle orientation is difficult in MC simulations, where the system continually shifts between configurations with different average tilt angles, which can be both positive and negative, we also show the average potential energy per particle $U_{\mathrm{avg}}$ (see Fig.\ \ref{fig:6}b). 
We typically find larger tilting and a higher potential energy when we approach, but not exceed, an integer ratio $C/L$ from below. This effect is tendentially stronger at larger area fractions. We observe these trends for both methods. 
The tilting allows the system to reduce the size of the gaps between the smectic layers. At larger area fractions, the lack of free space makes gaps between the smectic layers even more unfavorable, further favoring tilting over the formation of defects.

\section{\label{chap:conclusion}Conclusions}
We combined DFT and MC computer simulations to investigate the phase behavior of two-dimensional orientable hard rectangular particles both on a plane and on a cylindrical manifold for systems with and without aligning external fields. As a basis for our DFT calculations, we designed a new density functional that yields all liquid-crystalline phases observed in experiments with layers of hard rectangular particles \cite{NarayanMR2006, ZhaoHHRC2007, SanchezAM2016}. The resulting phase diagrams agree well with our particle-resolved simulations.

Depending on the aspect ratio and number density of the particles, we found stable isotropic, nematic, tetratic, and smectic phases in the flat and field-free case. Applying an aligning external field shifts the transition lines and enhances nematic order at the expense of tetratic order, which generates a binematic phase. For a cylindrical manifold, we observed in our DFT calculations both untilted and tilted smectic-like order around the cylinder. Additionally, the MC simulations showed both untilted and tilted columnar phases.

Future studies could generalize our DFT towards a dynamical density functional theory \cite{MarconiT1999, MarconiT2000, ArcherE2004, MarconiM2007, RexWL2007, EspanolL2009, WittkowskiL2011, GoddardNSYK2013}, which would provide insights into the nonequilibrium Brownian dynamics of such systems. It would also be interesting to consider other two-dimensional manifolds like cones and spheres or other particle interactions like those of ionic liquid crystals \cite{BartschBD2017} and magnetic nanorods \cite{SlyusarenkoCD2014}. Our results can be verified in experiments using sterically-stabilized rectangular colloidal particles \cite{LinCPSWLY2000,GalanisNLH2010, HermesVLVvODvB2011, ZvyagolskayaAB2011, SlyusarenkoCD2014, BesselingHKdNDDIvB2015, MuellerdlHRH2015,WalshM2016} or shaken granular particles \cite{CruzHidalgoZMP2010,HernandezNavarroIMST2012}.

\section*{Conflicts of interest}
There are no conflicts to declare.

\begin{acknowledgments}
We thank Axel Voigt for helpful discussions. R.W.\ and H.L.\ are funded by the Deutsche Forschungsgemeinschaft (DFG, German Research Foundation) -- WI 4170/3-1; LO 418/20-1. F.S.\ gratefully acknowledges funding from the Alexander von Humboldt foundation.
\end{acknowledgments}

\bibliographystyle{apsrev4-1}
\bibliography{refs}

\end{document}